%% file: main.tex
\documentclass[lettersize,journal]{IEEEtran}
\usepackage{amsmath,amsfonts}
\usepackage{algorithmic}
\usepackage{algorithm}
\usepackage{array}
\usepackage[caption=false,font=normalsize,labelfont=sf,textfont=sf]{subfig}
\usepackage{textcomp}
\usepackage{stfloats}
\usepackage{url}
\usepackage{verbatim}
\usepackage{graphicx}
\usepackage{cite}
\usepackage{xspace}
\usepackage{todonotes}
\usepackage{hyperref}
\usepackage{cleveref}
\usepackage{soul}
\usepackage{threeparttable}
\usepackage{wasysym}

\usepackage[utf8]{inputenc}

\input{macros}

\begin{document}

\title{Data Availability Sampling in Ethereum: Analysis of P2P Networking Requirements}

\author{\IEEEauthorblockN{Michał Król\IEEEauthorrefmark{3}, Onur Ascigil\IEEEauthorrefmark{2},  Sergi  Rene\IEEEauthorrefmark{1}, Etienne Rivière\IEEEauthorrefmark{4}, Matthieu Pigaglio\IEEEauthorrefmark{4} \\Kaleem Peeroo\IEEEauthorrefmark{3}, 
Vladimir Stankovic\IEEEauthorrefmark{3}, Ramin Sadre\IEEEauthorrefmark{4}, Felix Lange\IEEEauthorrefmark{5}}

 \vspace{+0.06in}
\IEEEauthorblockA{\IEEEauthorrefmark{1} University College London, 
}
\IEEEauthorblockA{\IEEEauthorrefmark{2} Lancaster University\\ 
}
\IEEEauthorblockA{\IEEEauthorrefmark{3} City, University of London,  
}
\IEEEauthorblockA{\IEEEauthorrefmark{4} ICTEAM, UCLouvain, Belgium \\ 
}
\IEEEauthorblockA{\IEEEauthorrefmark{5} Ethereum Foundation \\ 
}
\vspace{-0.2in}

}


\maketitle

\begin{abstract}
Despite their increasing popularity, blockchains still suffer from severe scalability limitations. Recently, Ethereum
proposed a novel approach to block validation based on Data Availability Sampling (DAS), that has the potential to improve its transaction per second rate by more than two orders of magnitude.
DAS should also significantly reduce per-transaction validation costs. 
At the same time, 
DAS introduces new communication patterns in the Ethereum peer-to-peer (P2P) network. These drastically increase the amount of exchanged data and impose stringent latency objectives. In this paper, we review the new requirements for P2P networking associated with DAS, discuss open challenges, and identify new research directions.
\end{abstract}


\begin{IEEEkeywords}
Ethereum, Scalability, Data Availability Sampling, Peer-to-peer, Networking.
\end{IEEEkeywords}

\input{./sections/1-intro}
\input{./sections/2-background}
\input{./sections/3-requirements}
\input{./sections/4-solutions}
\input{./sections/5-directions}
\input{./sections/6-conclusion}

\bibliographystyle{plain}
\bibliography{references}


\end{document}

%% file: macros.tex
\newboolean{showcomments}
\setboolean{showcomments}{false}

\newcommand\important[1]{\todo[inline]{\textbf{Important:} #1}}
\newcommand\etienne[1]{\todo[color=orange,inline]{\textbf{Etienne:} #1}}
\newcommand\felix[1]{\todo[color=blue!40,inline]{\textbf{Felix:} #1}}
\newcommand\michal[1]{\todo[color=green,inline]{\textbf{Michał:} #1}}
\newcommand\onur[1]{\todo[color=red,inline]{\textbf{Onur:} #1}}
\newcommand\sergi[1]{\todo[color=pink,inline]{\textbf{Sergi:} #1}}
\newcommand\ramin[1]{\todo[color=brown,inline]{\textbf{Ramin:} #1}}
\ifthenelse{\boolean{showcomments}} { }
{
\renewcommand\important[1]{}
\renewcommand\michal[1]{}
\renewcommand\onur[1]{}
\renewcommand\sergi[1]{}
\renewcommand\etienne[1]{}
\renewcommand\ramin[1]{}
\renewcommand\felix[1]{}
}

\ifthenelse{\boolean{showcomments}}
{ \newcommand{\mynote}[3]{
    \protect\fbox{\bfseries\sffamily\scriptsize#1}
    {\small\textsf{\emph{\color{#3}{#2}}}}}}
{ \newcommand{\mynote}[3]{}}

\newcommand{\er}[1]{\mynote{Etienne}{#1}{blue}}

\makeatletter
\def\BState{\State\hskip-\ALG@thistlm}
\makeatother

\newcommand{\eg}{\textit{e.g.}\@\xspace}

\newcommand{\ie}{\textit{i.e.}\@\xspace}

\newcolumntype{L}{l<{\hspace{1cm}}}
\newcolumntype{C}{c<{\hspace{1cm}}}
\newcolumntype{D}{c<{\hspace{0.3cm}}}

\newcommand*\feature[1]{\ifcase#1 -\or\LEFTcircle\or\CIRCLE\fi}

%% file: sections/1-intro.tex
\section{Introduction}

Ethereum is the second-largest blockchain currently in operation. 
Thanks to its support of smart contracts, it enables a wide range of applications from financial services and sharing-economy systems to Internet-of-Things, supply chains, and digital health~\cite{zile2018blockchain}. 

While the popularity of the Ethereum platform increases rapidly, its processing throughput remains very low, oscillating around 14 transactions per second (TPS). 
This low throughput is mainly a consequence of the consensus protocol~\cite{rouhani2017performance} that validators (\ie, platform maintainers) must run to agree on a common state. 
In each round of the consensus protocol, a chosen validator extends the blockchain by: \textit{(i)} creating a block containing the most profitable transactions; \textit{(ii)} linking the block to the previous one using a hash function, and \textit{(iii)} disseminating the block to other validators using a peer-to-peer (P2P) gossip protocol~\cite{gossipsub}. Every block must be delivered and \textit{verified} (\ie, checked for correctness) by all the validators in a timely manner. Having more honest validators in the system increases security but also causes additional communication and lowers performance, leading to a dilemma between throughput and decentralization.

An emerging solution to increase the scalability and throughput of Ethereum is the use of rollups~\cite{sguanci2021layer}. Rollups process their transactions off-chain and only periodically post small \emph{commitments} of their current state on-chain, where consensus is reached. This hybrid approach enables smaller, rollup-specific communities to process transactions at high speed while being eventually secured by the validators of the blockchain.


In contrast with transactions and smart contract executions registered on chain, commitment data posted by the rollups \textit{does not have to be verified} by Ethereum validators; it is sufficient to simply \emph{include} this data in new blocks.
Rollup participants can then verify commitment data using application-specific logic.
If this data is incorrect, rollup participants simply withdraw their funds and quit the application~\cite{sguanci2021layer}.

In a rollup implementation over the current, unmodified Ethereum deployment, commitment data is included in an opaque data blob that is an integral part of a block and competes for on-chain space with regular Ethereum transactions.
To limit the propagation time between the validator generating a new block and validators receiving and validating it, the size of a block is limited, and so is the size of this opaque data blob.
Thus, while rollups can significantly improve Ethereum throughput, this improvement is capped at a few thousand TPS due to block size limitations.

The inclusion of commitment data from rollups in blocks disseminated for validation is, in fact, a hindrance to reaching order-of-magnitude higher throughputs (e.g., 100,000 TPS).
Observing that commitment data must be available but is not checked for correctness by Ethereum validators, a new approach to block generation and validation emerged based on \emph{Data Availability Sampling} (DAS).
Unlike the previous approach, with DAS a validator only downloads in full the part of the block containing regular transactions, but not the opaque blob.
Instead, each validator collects sufficiently strong evidence that the opaque blob was indeed produced and made available by its producer.
The collection of this evidence is based on \textit{random sampling}, a process that becomes an integral part of the consensus: if the sampling process succeeds, validators consider the block correct.

DAS has the potential to maintain the high decentralization and security of Ethereum while massively increasing throughput.
However, its implementation presents significant challenges for the P2P network supporting the blockchain.
Currently, Ethereum uses simple gossiping techniques which are unsuitable for DAS data flows.
Indeed, the total amount of data generated per block increases from the current average of 90~KB up to 140~MB.
Even a resourceful block producer might not be able to \emph{directly} deliver the required samples to all network participants.
At the same time, relying on \emph{indirect} data dissemination is difficult.
First, each validator requests different samples, making gossip inefficient. 
Second, relying on anonymous network participants opens a new range of network-level attacks, threatening the correctness of the consensus protocol. 
Finally, the sampling process must be completed quickly (within 4~s for validators) as it determines whether the block is considered correct and can be referenced by future blocks in the chain; larger latencies significantly hinder progress. 

\begin{figure*}[t]
    \centering
    \includegraphics[width=0.8\linewidth]{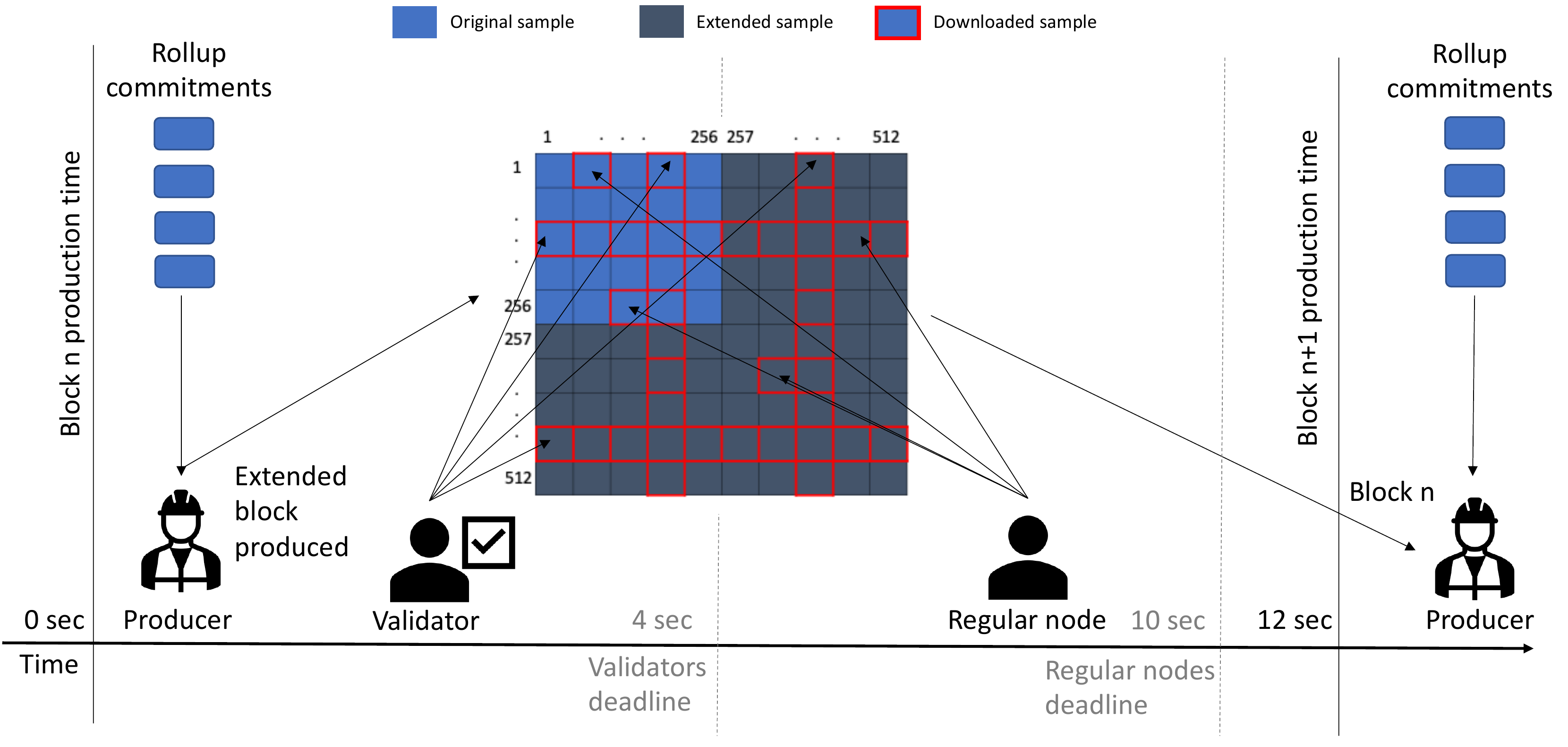}
    \vspace{-0.15in}
    \caption{Block data sampling. The producer gathers rollup commitments from the clients and creates an extended block. The block must be sampled by validators within 4~s from the block production by requesting 2 random rows and 2 random columns. Non-validator nodes sample the block within 10~s by requesting 75 random cells. After 12~s from the initial block production, the next block is created by a new block producer.}
    \vspace{-0.20in}
    \label{fig:ecosystem}
\end{figure*}

In this paper, we discuss design options for next-generation P2P networks that will be able to support Ethereum's rollup-centric DAS architecture at scale. We note that at the moment of writing, many details of the future DAS network are still under discussion. We base our analysis on emerging choices in the Ethereum development community, as they already give us a good sense of scale for this problem~\cite{neu2022data}.
In \Cref{sec:background}, we provide background on the new DAS approach. 
In \Cref{sec:requirements}, we determine concrete P2P network properties required to support the new model.
\Cref{sec:strawman} critically analyzes classical networking solutions and points out their shortcomings, \Cref{sec:solutions} lists new open challenges and research directions for the ecosystem, and \Cref{sec:conclusion} concludes the paper.

%% file: sections/2-background.tex
\section{Data Availability Sampling}
\label{sec:background}

Ethereum joins users that hold accounts and submit transactions, and validators that include transactions in blocks and maintain a full copy of the blockchain.
Ethereum recently migrated to a Proof-of-Stake (PoS) consensus protocol~\cite{merge-ethereum}.
Any node can now become a validator by staking at least 32~ETH, the Ethereum cryptocurrency. 
Time is divided into slots of 12 seconds and epochs of 32 slots. 
One validator is selected pseudo-randomly and based on stake to be a block producer in every slot\footnote{The Ethereum community recently proposed a \emph{Proposer-Builder Separation} mechanism (PBS) that delegates block production to dedicated \emph{builder} nodes. However, it does not change the requirements for the P2P network. Our discussion applies to both the current approach and this proposal.}.
This validator is responsible for creating a new block and sending it out to other nodes. 
Also in every slot, a committee of validators is chosen randomly, whose votes are used to determine the validity of the block being proposed.
The Ethereum design assumes that honest validators control a majority of the stake in the system and the majority of the voting power in each committee. 

The Data Availability Sampling (DAS) principle assumes that to every block is attached an opaque, binary blob (of up to 32 MB) holding rollup data.
Such a blob can be represented as a $256 \times 256$ matrix of 512~B cells. 
Both validators and regular Ethereum nodes (\eg, nodes that did not stake at least 32 ETH) do not verify the correctness of the blob (an operation that is specific to the different rollups) but only ensure that the blob was indeed released in full by the block producer.
To avoid downloading the entire blob for this purpose, the network relies on DAS (\ie, requesting a random subset of the data)~\cite{neu2022data}.
Theoretically, a malicious block producer could release all but a small part of the blob hoping that this will not be noticed by DAS.
Such behavior, called a \emph{data withholding attack}, threatens the security of rollups.
To prevent it, and as depicted in \Cref{fig:ecosystem}, the block is extended using a two-dimensional Reed-Solomon erasure code~\cite{wicker1999reed}. 
Each row and column doubles in size but can be reconstructed from any $50\%$ of its cells.
The extended block takes, therefore, the form of a $512 \times 512$ matrix, where each cell contains 512~B of data. 
Furthermore, each cell includes a 48~B cryptographic Kate-Zaverucha-Goldberg commitment (KZGC)~\cite{kate2010constant}, enabling the downloading node to make sure that the cell is a valid part of the extended blob. 
In total, the extended blob uses a total of 140~MB of data including 12~MB of KZGCs.
The original blob can now be reconstructed by acquiring at least 25\% of the extended blob cells, making a withholding attack much easier to detect.
Furthermore, KZGCs prevent a block producer from returning random data to sampling requests.


Depending on their role, nodes perform two types of data sampling:
\begin{itemize}
    \item \emph{Validators} randomly choose and attempt to download 2 rows and 2 columns of the extended blob ($2044$ cells in total).
    This \emph{validator sampling} is considered successful when at least 50\% (256 cells) of each column/row could be downloaded. 
    The remaining parts of the selected row/columns can be reconstructed using the erasure code.
    Each validator downloads at least $2046 \times 560\text{B} = 1.1~\text{MB}$ of data per slot.
    \item \emph{Regular nodes} randomly select and attempt to download 75 cells. 
    This \emph{regular sampling} is considered successful when all the selected cells can be downloaded.
    Each regular node downloads at least $75 \times 560\text{B} = 42~\text{KB}$ of data per slot.
    Note that validators perform regular sampling in addition to validator sampling.
\end{itemize}

If the sampling process fails, nodes consider the sampled block (and all the blocks that would later build upon it) invalid.
A malicious block producer may attempt to split the network by correctly responding to sampling requests from certain nodes while ignoring requests from others.
Such a behavior, called a \emph{network split attack}, results in honest nodes having a different view on the most recent valid block in the blockchain. As future blocks are built on top of the most recent block, the attack creates long-lasting forks, reducing both the efficiency and the security of the entire blockchain. The attack can be prevented by hiding the origin of sampling requests from block producers.

%% file: sections/3-requirements.tex
\section{What does DAS mean for the network?}
\label{sec:requirements}

Currently, the Ethereum network is formed of 500,000 validators and 2,000 regular nodes.
The target is, however, to support up to one million regular nodes participating in rollups.
All nodes first join a Distributed Hash Table (DHT) to discover other blockchain participants with whom they form an unstructured, mesh peer-to-peer (P2P) network for block dissemination~\cite{gossipsub,kiffer2021under}.
In the unmodified version of Ethereum, once a new block is created, its producer sends it to its direct peers that propagate the block further using epidemic dissemination.
Eventually, the block is delivered to all network participants.

However, such a solution cannot be adopted in the rollup-centric approach. Extended blobs are simply too big to be disseminated in full to the entire network during every 12~s slot. Furthermore, DAS assumes that each node downloads a small randomized subset of samples. Peers that are directly connected in the unstructured mesh are unlikely to request the same random samples, making the current epidemic dissemination intrinsically unsuitable.

\subsection{Objectives}

A P2P architecture supporting DAS should have the following functional and non-functional characteristics.

Functional requirements are threefold:
\begin{itemize}
    \item \textbf{Support two retrieval modes:} For the long-term security of the ecosystem, the solution should support both validator (rows and columns) and regular (random cells) sampling.
    \item \textbf{Openness:} Ethereum is an permissionless blockchain. The P2P network should remain open for anyone to join.
    \item \textbf{Request unlinkability:} To protect against network split attacks, for any two sampling requests, the block producer should not be able to tell whether they originate from the same sampling node.
\end{itemize}

In terms of non-functional requirements, we identify:
\begin{itemize}
    \item \textbf{Efficiency:} The network must be efficient in terms of the total amount of exchanged data by all its nodes. Higher overheads incur a higher monetary cost for the participants, threatening economic viability. They increase hardware requirements, hindering the decentralization of the network. An optimal solution transfers each requested sample exactly once. For the current network size, it means exchanging $500,000~\text{validators} \times 1.1~\text{MB} + 1,200~\text{nodes} \times 42~\text{KB} = 489~\text{GB}$ of data per slot.
    \item \textbf{Low dissemination latency:} The validators must have enough time to reach a consensus on the sampled block and start constructing the next block. Within a 12~s slot, the current consensus protocol requires that all requested samples be delivered within 4~s to the validators. Furthermore, to enable blockchain replication and reliable fork resolution the Ethereum community also suggests delivering samples to the regular nodes within 10~s~\cite{neu2022data}.
    \item \textbf{Low cost for the producer:} The block producer is incentivized to keep creating blocks only if the process generates a profit. The sampling procedure, being part of the block creation, should incur low monetary costs for the block producer and ensure that they do not exceed rewards from block creation.
    \item \textbf{Sybil attack resistance:} DAS is an internal part of the consensus protocol and must be secure. As the network is open for anyone to join, the solution should be resistant to malicious actors using Sybil identities (\ie, a large number of pseudonymous identities controlled by a single actor). 
\end{itemize}

%% file: sections/4-solutions.tex
\section{Classic Networking Approaches}
\label{sec:strawman}

We now review classical approaches to data distribution and distributed architectures, in light of the DAS network requirements.


\subsection{Centralized Approach}
\label{subsec:centralised}

The first approach would be for all nodes to retrieve samples directly from the block producer.
Joining the sampling process requires simply connecting to one of the producer's servers.
As each block might have a different producer, the IP address/the DNS domain must first reliably be advertised to the network.

This approach would require significant resources from block producers to deliver the large amount of sampling data requested by validators and regular nodes.
Using a well-provisioned cloud infrastructure, the cost for serving sampling requests for a single block would amount to roughly 25~USD\footnote{Taking as a reference the outbound data transfer cost for the Amazon AWS cloud \url{https://aws.amazon.com/ec2/pricing/on-demand/}, using a US-based datacenter}, or a monthly cost of around 5.75M~USD shared among all block producers.
This should account for the fact, however, that the block producers are chosen on a per-slot basis.
To be able to respond to such bandwidth demands in such a limited time, all potential block producers would have to set up permanent infrastructures that would need to be vastly over-provisioned, making cost figures significantly higher and severely impacting decentralization.
An alternative would be to rely on cloud-based Content Delivery Networks, but this would unacceptably put the progress of the chain in the hands of a few providers.
Another challenge would finally be to prevent Denial-of-Service (DoS) attacks targeting well-identified block producers.

Centralized sample delivery is conceptually simple and does not introduce intermediaries that may pose a security challenge.
It makes, however, the provision of \emph{request unlinkability} a challenge as requests are sent to the same location.
While validator sampling queries could be issued from 4 IP addresses (\ie, for 2 columns and 2 rows), regular sampling also requires that the 75 connections be indistinguishable from each other.
Using an anonymization network (e.g., Tor) may answer this problem but lead to a number of additional challenges regarding efficiency, robustness, and reliability.
While dedicated proxies may help, they would also become a critical part of the infrastructure and an easy target for attacks--simply magnifying the problem.

As a result, centralized approaches clearly go against the objectives of Ethereum in general, and DAS in particular, and should not further be considered.

\subsection{Unstructured P2P Network}
\label{subsec:gossip}

A peer-to-peer (P2P) network can assist in reducing the load on the block producer and increase decentralization.
The block producer can push samples to a few network participants that will propagate them further.
P2P networks establish an \emph{overlay} over the regular network, where nodes connect to a few (relatively to the size of the complete network) direct neighbors, or peers.

Unstructured P2P networks do not impose a particular, rigid structure on the P2P overlay network.
Nodes establish connections to each other either arbitrarily or following probabilistic choices leading to a globally random network~\cite{gossipsub,jelasity2007gossip}.

Data dissemination in unstructured P2P systems is typically achieved using gossip.
Gossip is efficient and robust for broadcast (\ie, sending the same message to everyone in the network) but requires additional measures for multicast (\ie, sending a message only to a specific subset).

Multicast is typical of publish/subscribe systems where nodes register to receive content posted to specific multicast groups, or topics.
Possible implementations of multicast include discovering the group during the dissemination, possibly leading to high overheads and the participation of many non-interested nodes in the dissemination.
Another approach is to pre-establish sub-networks linking only nodes from the same group~\cite{baldoni2007tera}.
This approach is, however, only beneficial when groups are stable and re-used multiple times.

While blockchain systems and Ethereum benefitted from the simplicity and effectiveness of gossip-based broadcast, DAS introduces specific, novel needs for multicast.
In a simple gossip scenario, if each node downloads and distributes only the samples it requires, the dissemination process never completes. On the other hand, making nodes participate in the distribution of additional samples (potentially up to a point where everyone downloads/distributes everything) is increasingly inefficient and requires additional resources when scaling up, reducing decentralization.

Pre-establishing dissemination structures is appealing but challenging due to the non-predictability of multicast groups.
Indeed, while supporting the validator sampling mode requires establishing $512 + 512=1,024$ dissemination structures, one for each column or row, the regular node mode increases this number by an additional $512 \times 512 = 262,144$, one per sample.
Additionally, each node randomly selects a different set of samples to download per slot, which means that the whole set of dissemination structures may need to be reconfigured every 12~s.
With frequent reconfigurations, the signaling overhead can bypass the volume of data being exchanged and the network might never reach a stable state. 

An alternative design could be for the producer to push samples to a \emph{subset} of nodes only.
Other participants would issue search queries to locate their content of interest.
However, this solution also introduces significant signaling overhead in terms of search queries.
Finding and downloading the relevant sample is unlikely to complete within the required 4 or 10 seconds. 

On the positive side, unstructured P2P solutions significantly reduce operational costs for the producer.
When each sample is delivered to a single node, the sampling process reduces the cost to 0,03\$ per block or 6,000\$ per month.
However, delivering each sample to a single node threatens the security of the approach as those nodes may turn malicious and simply refuse to propagate the sample further.
On the other hand, increasing the number of initial nodes also increases the monetary cost for the producer. 
This security risk can be reduced by a \textit{peer reputation mechanism}~\cite{gossipsub}.
It involves each node individually assessing the behavior of its direct neighbors and gradually replacing low-score peers. 
As a result, malicious nodes become isolated in the network and their impact is more limited.
Unfortunately, such a technique offers limited protection against scenarios where attackers behave correctly, increase their scores, and suddenly act maliciously. 

\er{I kept this paragraph but this would be a good candidate for removal, I don't think it adds much.}
To join the network, newcomers have to either already know some participants or use dedicated bootstrap nodes.
The latter, while used by almost all current P2P networks, is expensive for the bootstrap node operator and represents a single point of failure for the entire system. 
Unstructured P2P networks significantly reduce the possibility of data withholding attacks once the dissemination is underway.
The producer does not have a direct connection with all the nodes and cannot decide to ignore requests on a per-sample basis.
The producer can try to include its own Sybil nodes in the P2P network attempting to block certain nodes from accessing their samples.
However, due to the lack of structure and pseudorandom nature of data propagation such actions are not efficient and require a significant amount of resources increasing with the network size. 

\subsection{Distributed Hash Table (DHT)}
\label{subsec:dht}


A DHT is a structured overlay network where nodes and data objects (stored in the network) are both identified with unique keys (\ie, hash digests) drawn from a common hash space. DHTs allow efficient \textit{lookup} operations to locate node(s) in charge of a given key, \ie, nodes whose identifiers are the ``closest'' to a key are responsible for storing the data object
corresponding to that key.
DHTs use a \textit{distance metric} to measure the closeness of keys. 

A DHT node $s$ maintains a \textit{routing table} to store information, \ie, \emph{node ID} and reachability (IP address and port number), on $O(\log(n))$ peers arranged by their distance to $s$, where $n$ is the number of nodes in the DHT. Typically, a node's routing table provides a more detailed (\ie, fine-grained) view of the subset of the network with closer node IDs and a less detailed view of nodes with distant IDs. This property leads to efficient lookup operations that take a logarithmic number of steps (\ie, queried \emph{hops}) in the number of nodes in the network. Building on these efficient lookup operations, DHTs support a \textit{put} and \textit{get} API to store and retrieve data objects. 
There exist several DHT implementations~\cite{chord, maymounkov2002kademlia} that mostly differ in terms of distance metric, lookup procedure, routing table structure, and so on. Kademlia~\cite{maymounkov2002kademlia} is nowadays the most commonly deployed DHT protocol, including by Ethereum.

\textbf{DAS using a DHT:} Because of its efficient data storage and retrieval operations, using a DHT as a \textit{cache network}, where nodes collectively store and serve samples, is a natural choice for DAS. A DHT is a scalable network that can provide good load-balancing properties by assigning data objects uniformly to nodes. Ideally, a producer would only propagate a few copies of each cell of the block matrix to the DHT nodes responsible for storing them---\eg, using the hash of the meta-data of a cell (\eg, row, column numbers) as its key. 
The efficiency of DHT lookup ensures that the producer can store (\ie, put) samples in the DHT and that nodes retrieve (\ie, get) samples efficiently. 


An important threat to the operation of a DHT are Sybil identities. Because DHT nodes use self-generated identifiers (\ie, the hash of a public key), malicious actors can spawn multiple identities (even within one physical machine) to operate many DHT nodes. Sybils can then attempt to disrupt lookups and even launch various attacks as we discuss below. The impact of Sybils on the lookup procedure is rather different for \textit{iterative} and \textit{recursive} lookups.

With iterative lookups (as used by Kademlia) the querying node contacts in turn each intermediate node on the route to the destination.
Recursive lookups (as used by earlier DHTs~\cite{chord}) let each initially contacted node continue the query on behalf of the querier.
While iterative lookups increase the number of messages, they allow queries to better control the lookup and detect in particular if malicious nodes attempt to drop ongoing queries.
At the same time, iterative lookups are more difficult to anonymize as the querier directly communicates with all the nodes contacted during the operation. 


Another notable attack on DHTs is the \textit{eclipse attack}, where Sybils trick other nodes to populate their routing tables with only Sybils, thereby isolating those nodes and partitioning the network. Initially, each node populates its routing table through trusted bootstrap nodes that are discovered out-of-band. Over time, nodes discover new peers (\eg, ones who contact them) and store them in the routing tables subject to an \textit{admission policy}: the Kademlia implementation used both by Ethereum and by prominent web3 protocols, such as the decentralized storage system IPFS, limits the number of nodes from the same /24 subnet in the same bucket or routing table, making eclipsing more difficult---\ie, an adversary now has to deploy Sybils at multiple machines with diverse IP addresses to implement an attack.

S/Kademlia~\cite{pecori2016s} proposes a Proof-of-Work (PoW)-based node ID generation to make Sybil generation more resource-intensive. The main idea is to require at least a configurable number of 0’s at the end of the node IDs, which requires brute-force generation of potentially many public-private key pairs until an appropriate one is found. Another improvement by S/Kademlia is parallel iterative lookups that allow nodes to explore disjoint paths to obtain a higher probability of successful search results in the presence of Sybils.

DHT lookup overhead scales well with the number of participants but a single operation requires contacting a significant number of nodes (\eg, $\sim$50 for a 20,000-node network). Combined with a large number of samples in DAS and frequent slots in Ethereum, this can put a significant load on the network. The high bandwidth consumption may be problematic, especially for constrained, non-validator devices that are essential for the decentralization of the platform. Fortunately, the random nature of the sampling process allows us to distribute the load across the network and avoid hotspots that handle a disproportionate number of queries.

DHT lookups can potentially take a long time to complete with each contacted node by the querier being possibly far away in terms of geographical distance. Slow lookups can be problematic for DAS, which has stringent timing requirements for the completion of sampling operations. Recursive lookups can improve the latency (\ie, they require fewer round-trips), but they are also more difficult to secure, as discussed above.

%% file: sections/5-directions.tex
\section{Research Directions}
\label{sec:solutions}

A straightforward application of existing P2P (\ie, gossip and DHT) approaches fall short of satisfying the requirements of DAS in several aspects including security and delivery efficiency, while centralized approaches are unsuitable. In this section, we discuss possible directions to improve P2P approaches and make them suitable for DAS.

In terms of building a secure, Sybil-resistant P2P layer, one promising approach is to leverage the honest majority assumption of the consensus layer. In particular, the majority of the stake in the system is controlled by the honest parties as a security assumption of PoS. Therefore, an emerging approach is to use the honest majority assumption to build more robust P2P networks. 
Coretti \emph{et al.}~\cite{coretti2022generals} propose a peer sampling approach to choose neighbors in which nodes prefer staking nodes to build Sybil-resistant gossip networks. The resulting network has a highly connected backbone that Sybil nodes cannot partition as long as the honest majority assumption is not violated.


Currently, regular sampling fails when a single cell cannot be downloaded. It imposes high robustness requirements on the P2P networks that, due to its open and decentralized nature, might be challenging to fulfill. Designing \emph{$k$ out of $n$} sampling schemes might reduce those requirements and make P2P network deployments more practical. In such a scheme, regular nodes request $n > 75$ samples but still consider them successful when $k < n$ samples are downloaded within the time limit. Careful tuning $k$ and $n$ parameters can yield similar security guarantees to the current scheme while allowing the P2P network to tolerate sporadic node failures and malicious behavior. The scheme can be easily extended to improve the robustness of validator sampling. 

The ultimate weakness points of DHTs is the inability to protect specific places in the hash space. An attacker can ultimately generate thousands of peer IDs to strategically place themselves close to a specific key and hijack all the key-specific traffic. Storing/retrieving data from specific regions rather than from nodes closer to a specific key has the potential to remove this weakness. During an attack, while the data will be stored on/retrieved from malicious nodes, honest nodes are also guaranteed to fall within the region due to the uniform distribution of pseudorandom, legitimate peer IDs. As samples contain KZGC integrity proofs, sampling nodes can also easily filter out incorrect data. However, such an approach requires developing new, region-specific routing procedures in the DHT and redesigning the structure of its routing table.

When traversing the DHT, nodes ask their peers for information on how to access a specific part of the network. An attacker can attempt to hijack this process by returning only malicious nodes as a way for the node to progress. Currently, there are no constraints on the information returned by each participant that can be verified by the querying node. Carefully setting additional rules on the peers each node can return might significantly increase the security of DHT routing while maintaining its high efficiency.

One of the main problems of multicast DAS data dissemination is the necessity of frequent reconfiguration of a large number of pre-established groups that cause high signaling overhead. Making group membership more stable (\eg, across multiple slots) has the potential to make the multicast approach viable and much more efficient than broadcast-based dissemination. Alternatively, only a small number of nodes could migrate between fixed dissemination groups in every slot. However, such approaches require careful design to preserve the unpredictable nature of the samples requested by each node. 

The majority of P2P networks currently in use were developed about 20 years ago using hardware specifications that are no longer relevant. Current hardware is much more powerful in terms of memory and computational power and can support much higher bandwidths. For instance, storing routing information about 20,000 peers in a in Kademlia requires only $\sim$300~MB of memory. Adjusting P2P network parameters (\eg, DHT bucket sizes) can also significantly improve the performance of P2P networks to meet DAS requirements.

%% file: sections/6-conclusion.tex
\section{Conclusion}\label{sec:conclusion}
We revised the networking requirements for the next generation, rollup-centric Ethereum blockchain. While the current, classical approaches fall short of providing scalable network support for the new architecture, solving open research challenges has the potential to realize this vision. Supporting 100,000 transactions per second may push blockchains into mainstream use and fully realize their potential.